\documentclass{article}

\usepackage{spconf,amsmath,graphicx}

\usepackage{multirow}
\usepackage[font=scriptsize,caption=false,labelsep=space]{subfig}
\usepackage{mathrsfs}
\usepackage{graphicx,float,wrapfig,epstopdf,amsmath}
\usepackage[]{algorithmicx}
\usepackage{algpseudocode,algorithm}
\usepackage{balance}

\usepackage{amsmath}
\usepackage{amssymb}
\usepackage{amsthm}
\usepackage{graphicx}
\usepackage{epstopdf}
\usepackage{times}
\usepackage{textcomp,cite}
\usepackage{color}
\usepackage{url}

\usepackage{multirow}
\usepackage[table]{xcolor}
\usepackage{threeparttable}
\graphicspath{{./figures/}}

\DeclareMathOperator*{\minimize}{minimize} 

	\title{Federated Channel Learning for Intelligent Reflecting Surfaces With Fewer Pilot Signals}
	
\name{Ahmet~M.~Elbir$^{\dagger}$, Sinem Coleri$^+$ and Kumar Vijay Mishra$^\ddagger$ 
}
\address{	$^{\dagger}$Department of Electrical and Electronics Engineering, Duzce University, Duzce, Turkey\\
$+$Department of Electrical and Electronics Engineering, Ko\c{c} University, Istanbul, Turkey  \\
	$\ddagger$United States CCDC Army Research Laboratory, Adelphi, MD 20783 USA \\
	\thanks{S. C. acknowledges the support of Ford Otosan and the Scientific and Technological Research Council of Turkey EU CHIST-ERA grant 119E350.}\\
}

\begin{document}
\ninept		
		\maketitle
		
		\begin{abstract}
			Channel estimation is a critical task in intelligent reflecting surface (IRS)-assisted wireless systems due to the uncertainties imposed by environment dynamics and rapid changes in the IRS configuration. To deal with these uncertainties, deep learning (DL) approaches have been proposed. Previous works consider centralized learning (CL) approach for model training, which entails the collection of the whole training dataset from the users at the base station (BS), hence introducing huge transmission overhead for data collection. To address this challenge, this paper proposes a federated learning (FL) framework to jointly estimate both direct and cascaded channels in IRS-assisted wireless systems. We design a single convolutional neural network trained on the local datasets of the users without sending them to the BS. We show that the proposed FL-based channel estimation approach requires approximately $60\%$ fewer pilot signals and it  exhibits  12 times lower transmission overhead than CL, while maintaining satisfactory performance close to CL. In addition, it provides lower estimation error than the state-of-the-art DL-based schemes.
		\end{abstract}

		\section{Introduction}
		\label{sec:Introduciton}
		
		Intelligent reflecting surfaces (IRSs) have been attracted a great research interest due to the lower hardware complexity, physical size and cost of conventional large arrays~\cite{irs_survey1}. An IRS is an electromagnetic two-dimensional surface that is composed of large number of passive reconfigurable meta-material elements, which reflect the incoming signal by introducing a pre-determined phase shift. This phase shift is controlled via external signals by the base station (BS) through a backhaul control link. As a result, the incoming signal from the BS can be manipulated in real-time, thereby, reflecting the received signal toward the users. Hence, the usage of IRS enhances the signal energy received by distant users and expands the coverage of the BS~\cite{lis_channelEst_3}. 
		
		The system configuration of the IRS-assisted systems, i.e., designing reflect beamformers, strongly relies on the knowledge of the channel information. In fact, the IRS-assisted systems include multiple communications links, i.e., a direct channel from BS to users and a cascaded channel from BS to users through IRS. This makes the IRS scenario even more challenging than the conventional massive MIMO (multiple-input multiple-output) systems. Furthermore, the wireless channel is dynamic and uncertain because of changing IRS configurations. Consequently, there exists an inherit uncertainty stemming from the IRS configuration and the channel dynamics. These characteristics of IRS make the system design very challenging~\cite{elbir2020IRS_DL_survey,irs_deepDenoisingNN_CE}.
		
		To address the aforementioned uncertainties and non-linearities imposed by channel equalization, hardware impairments, and sub-optimality of high-dimensional problems, model-free techniques have become common in wireless communications~\cite{elbir2020IRS_DL_survey}. In this context, deep learning (DL) is particularly powerful in extracting the features from the raw data by constructing a model-free data mapping. DL-based channel estimation for IRS-assisted wireless systems is considered in~\cite{elbir_LIS,irs_deepDenoisingNN_CE}. Specifically, \cite{elbir_LIS}   devise a twin convolutional neural network (CNN), which takes the least squares (LS) estimates of the direct and cascaded channels as input to estimate the channels. A deep denoising neural network (DDNN) approach is proposed in~\cite{irs_deepDenoisingNN_CE}, where the input of the DDNN is a compressive channel estimate that is obtained via orthogonal matching pursuit (OMP) algorithm from the receive antenna array measurements. These DL techniques employ a centralized learning (CL) scheme, wherein the learning model is trained at a parameter server (PS) (possible at the BS) by collecting the training data from the users. The transmission of the training datasets from the users to the PS introduces huge overhead due to the large size of the datasets. This overhead can be reduced by decentralized techniques, such as federated learning (FL), by bringing the learning task to the edge level, e.g., the mobile users~\cite{spm_federatedLearning,fl_By_Google,elbir2021FL4PHY}. In FL, instead of transmitting the whole  dataset to the PS, each user computes the local model updates on its local dataset and only transmits the model updates (gradients) to the PS. Due to its recent success on wireless sensor networks~\cite{FL_Gunduz,fl_IoT}, UAV (unmanned aerial vehicle) networks~\cite{FL_Bennis2} and  vehicular networks~\cite{elbir2020federated}, FL has been regarded as a promising tool to exhibit a communication-efficient as well as privacy-preserving learning scheme since it does not involve raw data transmission. Recently, FL has been applied for beamformer design problems in massive MIMO~\cite{elbir2020FL_HB} and IRS-assisted systems~\cite{elbir2020_FL_CE}. In~\cite{elbir2020_FL_CE}, FL is applied for channel estimation in IRS-assisted systems, where the proposed approach requires pilot signals at least the number of BS antennas. The present work eliminates such requirement and provides satisfactory channel estimation performance close to CL. 

		%
		%

		In this paper, we propose an FL approach for channel estimation in IRS-assisted wireless systems. We design a single CNN to jointly estimate the direct and cascaded channels in the presence of insufficient pilot signals. During model training, each user computes the model updates by using the local datasets (i.e., input of received pilots and label of channel data) and transmits them to the BS where the model updates are aggregated and sent back to the users. The proposed FL approach is advantageous since it provides decentralized learning, which	significantly reduces the transmission overhead compared to the CL-based techniques while maintaining satisfactory channel estimation performance close to CL. Furthermore, the performance of the proposed method can be attributed to the fact that the designed learning model constructs a non-linear mapping from the insufficient number of received pilots to the channel matrix entries. Therefore, the proposed approach is also effective since it does not involve a matrix inversion operation, which may be required in model-based techniques in massive MIMO applications. To the best of our knowledge, this is the first FL-based method for channel estimation with fewer pilot signals.
		

		\section{Signal Model}
		\label{sec:FL_IRS}
		We consider the downlink channel estimation for IRS-assisted massive MIMO systems, where the BS has $M$ antennas to serve $K$ single-antenna users with the assistance of IRS, which is composed of $L$ reflective elements. The incoming signal from the BS is reflected from the IRS, where each IRS element introduces a phase shift $\varphi_l$, for $l=1,\dots, L$. This phase shift can be adjusted through the PIN (positive-intrinsic-negative) diodes, which are controlled by the IRS-controller connected to the BS over the backhaul link~\cite{lis_channelEst_3,lis_channelEst_4}. As a result, IRS allows the users receive the signal transmitted from the BS when they are distant from the BS or there is a blockage among them. Assume that the BS transmits the $k$-th user the data symbol $s_k\in\mathbb{C}$ by using a baseband precoder $\mathbf{F}= [\mathbf{f}_1,\dots,\mathbf{f}_K]\in\mathbb{C}^{M\times K}$ . Hence, the downlink $M\times 1$ transmitted signal becomes $	\overline{\mathbf{s}} = \sum_{k=1}^{K} \sqrt{\gamma_k} \bar{\mathbf{f}}_ks_k,$	where {\color{black} $\bar{\mathbf{f}}_k = \frac{\mathbf{f}_k}{||\mathbf{f}_k||_2}$ and} $\gamma_k$ denotes the allocated power at the $k$-th user. The transmitted signal is received from the $k$-th user with two components, one of which is through the direct path from the BS and another one is through the IRS. The received signal at the $k$-th user can be given by
		\begin{align}
		\label{receivedSignal1}
		y_k = \big(\mathbf{h}_{\mathrm{BS},k}^\textsf{H} + \mathbf{h}_{\mathrm{IRS},k}^\textsf{H} \boldsymbol{\Psi}^\textsf{H} \mathbf{H}^\textsf{H}  \big) \overline{\mathbf{s}} + n_k,
		\end{align}
		where $n_{k}\in\mathbb{C}$ denotes the spatially and temporarily white Gaussian noise and 
		$\boldsymbol{\Psi} = \mathrm{diag}\{\boldsymbol{\psi}\}\in\mathbb{C}^{L\times L}$ and $\boldsymbol{\psi} = [\psi_1, \dots, \psi_{L}]^\textsf{T}\in \mathbb{C}^{L}$ is the reflecting beamformer vector, whose $l$-th entry is $\psi_l = a_n e^{j \varphi_l} $, where $a_l\in \{0,1\}$ denotes the on/off stage of the $l$-th element of the IRS and $\varphi_l \in [0, 2\pi]$ is the phase shift introduced by the IRS. In practice, the IRS elements cannot be perfectly turned on/off, hence, they can be modeled as $		a_n=\left\{\begin{array}{cc}
		1 - \epsilon_1 & \mathrm{On}\\
		0 + \epsilon_0 & \mathrm{Off}
		\end{array}\right.,$
		for small $\epsilon_1,\epsilon_0 \geq0$, which represents the insertion loss of the reflecting elements~\cite{elbir_LIS,lis_onoff_ICASSP}. 
		$\mathbf{H}\in \mathbb{C}^{M \times L}$ is the channel between the BS and the IRS and it can be defined as $	\mathbf{H} = \sqrt{\frac{ML }{N_\mathrm{r}}} \sum_{n=1}^{ N_\mathrm{r}} \alpha_{n} \mathbf{a}_\mathrm{BS}( \phi_{n}^{\mathrm{BS}})\mathbf{a}_\mathrm{IRS}( {\phi}_{n}^{\mathrm{IRS}})^\textsf{H},$
	where $N_\mathrm{r}$ and $\alpha_{l}^{\mathrm{IRS}}$ are the number of received paths and the complex gain respectively. $\mathbf{a}_\mathrm{BS}( \phi_{n}^{\mathrm{BS}})\in \mathbb{C}^{M}$ and $\mathbf{a}_\mathrm{IRS}( {\phi}_n^\mathrm{IRS})\in \mathbb{C}^{L}$ are the steering vectors corresponding to the BS and IRS with the angle-of-arrival/departure (AoA/AoD) angles $\phi_{n}^{\mathrm{BS}},\phi_{n}^{\mathrm{IRS}}$, respectively. $\mathbf{h}_{\mathrm{BS},k}$ $\in\mathbb{C}^{M}$ and $\mathbf{h}_{\mathrm{IRS},k}\in \mathbb{C}^L$ represent the direct channels for BS-user and IRS-user links, and they can be defined respectively as 
		\begin{align}
		    \mathbf{h}_{\mathrm{BS},k} = \sqrt{\frac{M }{N_\mathrm{r}^\mathrm{BS}  }} \sum_{n=1}^{ N_\mathrm{r}^\mathrm{BS}} \alpha_{k,n}^{\mathrm{BS}} \mathbf{a}_\mathrm{BS}( \varphi_{k,n}^{\mathrm{BS}}),
		\end{align} and
		\begin{align}
		\mathbf{h}_{\mathrm{IRS},k} = \sqrt{\frac{L }{N_\mathrm{r}^{\mathrm{IRS}}  }} \sum_{l=1}^{ N_\mathrm{IRS}} \alpha_{k,n}^{\mathrm{IRS}} \mathbf{a}_\mathrm{IRS}( \varphi_{k,n}^{\mathrm{IRS}}).
		\end{align} Here, 	 $N_\mathrm{r}^\mathrm{BS}$, $\alpha_{k,n}^{\mathrm{BS}}$ and $\mathbf{a}_\mathrm{BS}(\varphi_{k,n}^{\mathrm{BS}})$ ($N_\mathrm{r}^\mathrm{IRS}$, $\alpha_{k,n}^{\mathrm{IRS}}$, $\mathbf{a}_\mathrm{IRS}(\varphi_{k,n}^{\mathrm{IRS}})$) are the number of paths, complex gain and the steering vector with the AoA angle $\varphi_{k,n}^{\mathrm{BS}}$ ($\varphi_{k,n}^{\mathrm{IRS}}$) for the BS-user (IRS-user) communication link, respectively. 
		
		Let $\boldsymbol{\Lambda}_k =  \mathrm{diag}\{ \mathbf{h}_{\mathrm{IRS},k}\}\in \mathbb{C}^{L\times L}$, then the cascaded channel between the BS and the $k$-th user becomes $\mathbf{G}_k = \mathbf{H} \boldsymbol{\Lambda}_k\in \mathbb{C}^{M\times L}$ and (\ref{receivedSignal1}) can be rewritten as
		\begin{align}
		y_k = \big(\mathbf{h}_{\mathrm{BS},k}^\textsf{H} + \boldsymbol{\psi}^\textsf{H} \mathbf{G}_k^\textsf{H}  \big) \overline{\mathbf{s}} + n_k.
		\end{align}
		
		Our aim in this work is to estimate the direct and cascaded channels via FL. In the following, we first introduce how input and output data are designed for the proposed learning model, then discuss how FL-based training is performed.

		\section{Channel  Acquisition With Fewer Pilots In IRS-Assisted Wireless Systems}
		The proposed channel estimation scheme has two stages to separately collect the received data for direct and cascaded channels, $\mathbf{h}_{\mathrm{BS},k}$ and $\mathbf{G}_k$, respectively. 
		In order to estimate the channels $\{\mathbf{h}_{\mathrm{BS},k}, $ $\mathbf{G}_k\}$, $k = 1,\dots,K$, conventional model-based~\cite{lis_channelEst_3,lis_channelEst_4} and model-free~\cite{deepCNN_ChannelEstimation,elbir_LIS,elbir2020_FL_CE} techniques require at least $\bar{M}\geq M$ orthogonal pilot signals, which can be represented as $\mathbf{S} = [\mathbf{s}_1,\dots, \mathbf{s}_M]\in\mathbb{C}^{M\times M}$ whose columns are orthogonal to each other, i.e., $||\mathbf{s}_{m_1}^\textsf{H}\mathbf{s}_{m_2} ||_2^2 =0$ for $m_1 \neq m_2$, $m_1,m_2=1,\dots,M$. In the presence of insufficient pilot signals, i.e., $\bar{M}< M$, using the incomplete pilot matrix $\bar{\mathbf{S}}\in \mathbb{C}^{M\times \bar{M}}$, the received signal at the $k$-th user becomes
		\begin{align}
		\label{receivedIRS1}
		\mathbf{y}_k = (\mathbf{h}_{\mathrm{BS},k}^\textsf{H} + \boldsymbol{\psi}^\textsf{H}\mathbf{G}_k^\textsf{H}) \overline{\mathbf{S}} + \mathbf{n}_k,
		\end{align}
		where $\mathbf{y}_k = [y_{1,k},\dots, y_{\bar{M},k}]$ and $\mathbf{n}_k = [n_{1,k},\dots, n_{\bar{M},k}]$ are $1\times \bar{M}$ row vectors. We intend to collect the received data with insufficient number of pilots, then construct a learning model which maps the received data to the channel data.
		
		In the first stage, the received data with respect to $\mathbf{h}_{\mathrm{BS},k}$ can be collected when all of the IRS elements are turned off, i.e., $a_l = 0$, for $l=1,\dots,L$. Then, the $1\times \bar{M}$ received signal at the $k$-th user is
		\begin{align}
		\label{esthdirect}
		\mathbf{y}_{\mathrm{D},k} = \mathbf{h}_{\mathrm{BS},k}^\textsf{H}\overline{\mathbf{S}} + \mathbf{n}_{\mathrm{BS},k}.
		\end{align}
		The conventional model-based methods, such as least squares (LS) and minimum mean-squared-error (MMSE), fail to obtain $\mathbf{h}_{\mathrm{BS},k}$ from $\mathbf{y}_{\mathrm{BS},k}$ due to the lack of sufficient pilot data. Here, we leverage DL to estimate $\mathbf{h}_{\mathrm{BS},k}$ by selecting $\mathbf{y}_{\mathrm{B},k}$ as input to the CNN.

		Next, we consider the cascaded channel estimation. We assume that each IRS element is turned on one by one while all the other elements are turned off. This is done by the BS requesting the IRS via a micro-controller device in the backhaul link so that a single IRS element is turned on at a time. Then, the reflecting beamformer vector at the $l$-th frame becomes $\boldsymbol{\psi}^{(l)} = [0,\dots, 0, \psi_l,0,\dots, 0]^\textsf{T}$, where $a_l = \{0: \tilde{l} = 1,\dots L, \tilde{l} \neq l \}$ and the received signal becomes
		\begin{align}
		\label{esthcascaded}
		\mathbf{y}_{\mathrm{C},k}^{(l)} = (\mathbf{h}_{\mathrm{BS},k}^\textsf{H} +{\mathbf{g}_{k}^{(l)}}^\textsf{H} ) \overline{\mathbf{S}} + \mathbf{n}_k,
		\end{align}
		where $\mathbf{g}_k^{(l)} \in \mathbb{C}^{M}$ is the $l$-th column of $\mathbf{G}_k$, i.e., $	\mathbf{g}_k^{(l)} = \mathbf{G}_k \boldsymbol{\psi}^{(l)},$ where  $\psi_l = 1$. Using the estimate of $\mathbf{h}_{\mathrm{BS},k}$ from (\ref{esthdirect}), (\ref{esthcascaded}) can be solved for $\mathbf{g}_k^{(l)}$, $l = 1,\dots, L$, and the cascaded channel $\mathbf{G}_k$ can be estimated. Then, the  received data for $l = 1\dots, L$ can be collected as
		\begin{align}
		\mathbf{Y}_{\mathrm{C},k} =\left[  
		\mathbf{y}_{\mathrm{C},k}^{(1)^\textsf{T}},\dots, 
		\mathbf{y}_{\mathrm{C},k}^{(L)^\textsf{T}}\right]^\textsf{T}\in \mathbb{C}^{L\times \bar{M}}.
		\end{align}
		
		In order to train the CNN for IRS-assisted massive MIMO scenario, we select  the input-output data pair as $\{\mathbf{y}_{\mathrm{D},k},\mathbf{h}_{\mathrm{BS},k} \}$ and $\{\mathbf{Y}_{\mathrm{C},k},\mathbf{G}_{k} \}$ for direct and cascaded channels, respectively. To jointly learn both channels, a single input is constructed to train a single CNN as  $ \boldsymbol{\Upsilon}_k = \left[\begin{array}{c}
		\mathbf{y}_{\mathrm{D},k}\\
		\mathbf{Y}_{\mathrm{C},k}
		\end{array}\right] \in \mathbb{C}^{(L + 1)\times \bar{M}}$. We design the input of the CNN as three \textit{channel} of the input data, i.e., $[{\mathcal{X}}_k]_1 = \operatorname{Re}\{ \boldsymbol{\Upsilon}_k\}$ and $[{\mathcal{X}}_k]_2 = \operatorname{Im}\{ \boldsymbol{\Upsilon}_k\}$, $[{\mathcal{X}}_k]_3 = \angle\{ \boldsymbol{\Upsilon}_k\}$, respectively. We can define the output data as $\boldsymbol{\Sigma}_k = \left[\mathbf{h}_{\mathrm{BS},k}, \mathbf{G}_{k}\right] \in \mathbb{C}^{\bar{M}\times (L+1)}$, hence, the output label can be given by a $2{M}(L+1)\times 1$ real-valued vector as 
		\begin{align}
		{\mathcal{Y}}_k = \left[ \mathrm{vec}\{\operatorname{Re} \{\boldsymbol{\Sigma}_k\}\}^\textsf{T},  \mathrm{vec}\{\operatorname{Im} \{\boldsymbol{\Sigma}_k\}\}^\textsf{T}   \right]^\textsf{T}.
		\end{align}	Consequently, the sizes of ${\mathcal{X}}_k$ and ${\mathcal{Y}}_k$ are $(L + 1)\times \bar{M}\times 3$ and $2{M} (L+1)\times 1$, respectively. Once the CNN is trained, the received signals $\mathbf{y}_{\mathrm{D},k}$ and $\mathbf{Y}_{\mathrm{C},k}$ can be collected and fed to the CNN to estimate the channels. We discuss the FL-based model training in the following section.

		\section{FL-based Model Training}
		Let $\mathcal{D}_k$ be the local dataset of the $k$-user, in which the $i$-th element is given by $\mathcal{D}_i = (\mathcal{X}_k^{(i)},\mathcal{Y}_k^{(i)})$, where  $\mathcal{X}_k^{(i)}$ and $\mathcal{Y}_k^{(i)}$ denote the input and output  for $i =1,\dots,\textsf{D}_k$ and $\textsf{D}_k = |\mathcal{D}_k|$ ($\textsf{D} = \sum_k \textsf{D}_k $) is the size of the local dataset.	We begin by introducing the training concept in conventional CL-based training, then develop FL-based model training. 
		
		In CL-based model training for channel estimation~\cite{deepCNN_ChannelEstimation,elbir2019online,elbir_LIS}, the training of the global CNN is performed by collecting the local datasets $\mathcal{D}_k, k = 1,\dots,K$ from the users. Once the BS has collected the whole dataset $\mathcal{D} = \bigcup_{k} \mathcal{D}_k $, the training is performed by solving the following problem
		\begin{align}
		\label{lossML}
		\minimize_{\boldsymbol{\theta}}    
		\mathcal{F}(\boldsymbol{\theta}) =  \frac{1}{\textsf{D}} \sum_{i = 1}^{\textsf{D}}\mathcal{L}(f( \mathcal{X}^{(i)}|\boldsymbol{\theta}),\mathcal{Y}^{(i)}  )  ,
		\end{align}
		where $\boldsymbol{\theta}\in\mathbb{R}^P$ denotes the learnable parameters and $\mathcal{L}(\cdot)$ is the loss function defined by the learning model as
		\begin{align}
			\mathcal{L}(f(\mathcal{X}^{(i)}|\boldsymbol{\theta}),\mathcal{Y}^{(i)}) =  \| f( \mathcal{X}^{(i)}|\boldsymbol{\theta}) - \mathcal{Y}^{(i)}  \|_\mathcal{F}^2,
		\end{align}
		which is the MSE between the label data $\mathcal{Y}^{(i)}$ and the prediction of the CNN, $f( \mathcal{X}^{(i)}|\boldsymbol{\theta})$ for the whole dataset, i.e., $i = 1,\dots, \textsf{D}$.  The minimization of the empirical loss $	\mathcal{F}(\boldsymbol{\theta})$ is achieved via gradient descent (GD) by updating the model parameters $\boldsymbol{\theta}_t$ at iteration $t$ as $\boldsymbol{\theta}_{t+1} = \boldsymbol{\theta}_t - \eta_t \mathbf{g}(\boldsymbol{\theta}_t),$
		where $\eta_t$ is the learning rate and 
		\begin{align}
		\mathbf{g}(\boldsymbol{\theta}_t)= \nabla_{\boldsymbol{\theta}} \mathcal{F}(\boldsymbol{\theta}_t) =\frac{1}{\textsf{D}} \sum_{i=1}^{\textsf{D}}\nabla_{\boldsymbol{\theta}} \mathcal{L} (f ( \mathcal{X}^{(i)}|\boldsymbol{\theta}_t), \mathcal{Y}^{(i)}),
		\end{align}
		denotes the \emph{full} or \emph{batch} gradient vector in  $\mathbb{R}^P$~\cite{FL_QSGD,fl_convergenceOnNIIDData}. For large datasets, it is computationally inefficient to implement GD, which motivates the use of stochastic GD (SGD), where $\mathcal{D}$ is partitioned into $M_B$ mini-batches as $\mathcal{D} = \bigcup_{m\in \mathcal{M}_B} \mathcal{D}_m$, for $\mathcal{M}_B = \{1,\dots, M_B\}$. Then,   $\boldsymbol{\theta}_t$ is updated by $\boldsymbol{\theta}_{t+1} = \boldsymbol{\theta}_t - \eta_t \mathbf{g}_{\mathcal{M}_B}(\boldsymbol{\theta}_t),$
		where ${\mathbf{g}}_{\mathcal{M}_B}(\boldsymbol{\theta}_t) =\frac{1}{M_B} \sum_{m=1}^{M_B} \mathbf{g}_m(\boldsymbol{\theta}_t) $ includes the contribution of gradients computed over $\{\mathcal{D}_m\}_{m\in \mathcal{M}_B}$ as $	\mathbf{g}_m(\boldsymbol{\theta}_t) = \frac{1}{\textsf{D}_m} \sum_{i=1}^{\textsf{D}_m}\nabla_{\boldsymbol{\theta}} \mathcal{L} (f ( \mathcal{X}_m^{(i)}|\boldsymbol{\theta}_t), \mathcal{Y}_m^{(i)}) ,$
		where ${\textsf{D}_m} = |\mathcal{D}_m|$ is the mini-batch size and ${\mathbf{g}}_{\mathcal{M}_B}(\boldsymbol{\theta}_t)$ satisfies $\mathbb{E}\{ {\mathbf{g}}_{\mathcal{M}_B}(\boldsymbol{\theta}_t) \} = \nabla_{\boldsymbol{\theta}} \mathcal{F}(\boldsymbol{\theta}_t)$. Therefore, SDG provides the minimization of the empirical loss by partitioning the dataset into $M_B$ mini-batches and accelerates the learning process, which is known as mini-batch learning.
		
		In FL, the training dataset $\mathcal{D}$ is partitioned into small portions, i.e., $\mathcal{D}_k$, $k = 1,\dots,K$, however they are available at the users and not transmitted to the BS. Hence, given initial model $\boldsymbol{\theta}_{t-1}$, the $k$-th user aims to solve the following local problem at iteration $t$, i.e.,
		\begin{align}
		\label{lossFL}
		\minimize_{\boldsymbol{\theta}}    
		\frac{1}{\textsf{D}_k} \sum_{i = 1}^{\textsf{D}_k}\mathcal{L}(f( \mathcal{X}_k^{(i)}|\boldsymbol{\theta}_{t-1}),\mathcal{Y}_k^{(i)}  ),
		\end{align}
		with the use of the local gradient $\mathbf{g}_k(\boldsymbol{\theta}_t)$. The $k$-th user transmits $\mathbf{g}_k(\boldsymbol{\theta}_t)$ to the BS, which receives $\tilde{\mathbf{g}}_k(\boldsymbol{\theta}_t) = \mathbf{g}_k(\boldsymbol{\theta}_t) + \mathbf{e}_{k,t},$
		where $\mathbf{e}_{k,t}\in \mathbb{R}^P$ denotes the noise term added onto $\mathbf{g}_k(\boldsymbol{\theta}_t)$ at the $t$-th iteration. Without loss of generality, we assume that $\mathbf{e}_{k,t}$ obeys normal distribution, i.e., $\mathbf{e}_{k,t}\sim \mathcal{N}(0,\sigma_{\boldsymbol{\theta}}^2\mathbf{I}_P)$, for $\mathbf{I}_P$ being $P\times P$ identity matrix with variance $\sigma_{\boldsymbol{\theta}}^2$ and the signal-to-noise-ratio (SNR) in gradient transmission is given by  $\mathrm{SNR}_{\boldsymbol{\theta}} = 20\log_{10}\frac{||\mathbf{g}_k(\boldsymbol{\theta}_t)||_2^2}{\sigma_{\boldsymbol{\theta}}^2  }  $. Once the gradient data from all users are collected, the BS finally incorporates $\tilde{\mathbf{g}}_k(\boldsymbol{\theta}_t)$ for $k = 1,\dots,K$ to update $\boldsymbol{\theta}_t$ as $\boldsymbol{\theta}_{t+1} = \boldsymbol{\theta}_t - \eta_t  \frac{1}{K} \sum_{k=1}^{K} \tilde{\mathbf{g}}_k(\boldsymbol{\theta}_t),$
		where $	\tilde{\mathbf{g}}_k(\boldsymbol{\theta}_t) = \frac{1}{\textsf{D}_k} \sum_{i=1}^{\textsf{D}_k}\nabla_{\boldsymbol{\theta}} \mathcal{L} (f ( \mathcal{X}_k^{(i)}|\boldsymbol{\theta}_t), $ $\mathcal{Y}_k^{(i)})  ).$
		 After model aggregation, the BS returns the updated model parameters $\boldsymbol{\theta}_{t+1}$ to the users, and the received model parameter at the $k$-th user can be written as $\tilde{\boldsymbol{\theta}}_{t+1} = \boldsymbol{\theta}_{t+1} + \mathbf{e}_{k,t+1},$
		which will be used for the computation of the gradients in the next iteration.

		\subsection{Neural Network Architecture}
		\label{sec:NNarch}
		We design a single CNN trained on the local datasets of the users via FL. The proposed network architecture is comprised of $10$ layers. The first layer is the input layer, which accepts the input data of size $(L + 1)\times \bar{M}\times 3$.  The $\{2,4,6\}$-th layers are the convolutional layers with $N_\mathrm{SF} = 128$ filters, each of which employs a $3\times 3$ kernel for 2-D spatial feature extraction. The $\{3,5,7\}$-th layers are the normalization layers. The eighth layer is a fully connected layer with $N_\mathrm{FCL}=1024$ units, whose main purpose is to provide feature mapping. The ninth layer is a dropout layer with $\kappa=1/2$ probability. The dropout layer applies an $N_\mathrm{FCL}\times 1$ mask on the weights of the fully connected layer, whose elements are uniform randomly selected from $\{0,1\}$. As a result, at each iteration of FL training, randomly selected different set of weights in the fully connected layer is updated. Thus, the use of dropout layer reduces the size of $\boldsymbol{\theta}_t$ and $\mathbf{g}_k(\boldsymbol{\theta}_t)$, thereby, reducing model transmission overhead. Finally, the last layer is output regression layer, yielding the output channel estimate of size  $2M (L+1)\times 1$. The CNN architecture is similar to the one in~\cite{elbir2020_FL_CE} with the main difference being that the current model accepts $(L + 1)\times \bar{M}\times 3$ input with fewer pilots instead of $(L + 1)\times {M}\times 3$ as in\cite{elbir2020_FL_CE}.

		\subsection{Transmission Overhead}
		\label{sec:DataComp}
		Transmission overhead can be defined as the size of the transmitted data during model training~\cite{elbir2020_FL_CE,FL_gunduz_fading,FL_Gunduz,fl_By_Google}. Let $\mathcal{T}_\mathrm{FL}$ and $\mathcal{T}_\mathrm{CL}$ denote the transmission overhead of FL and CL, respectively. Then, we can define $\mathcal{T}_\mathrm{CL}$ as $	\mathcal{T}_\mathrm{CL} = 
	 \big(3\bar{M}(L + 1) + 2M (L+ 1)\big) \textsf{D},$
		which includes the number of symbols in the uplink transmission of the training dataset from the users to the BS. In contrast, the transmission overhead of FL includes the transmission of $\mathbf{g}_k(\boldsymbol{\theta}_t)$ and $\boldsymbol{\theta}_t$ in uplink and downlink communication for $t = 1,\dots,T$, respectively. Compared to~\cite{elbir2020_FL_CE}, the first term on the right hand side of $\mathcal{T}_\mathrm{CL}$ is computed with $\bar{M}$ instead of $M$. Finally, $\mathcal{T}_\mathrm{FL}$ is given by
		\begin{align}
		\mathcal{T}_\mathrm{FL} = 
		2PTK.
		\end{align} We can see that the dominant terms are $\textsf{D}$ and $P$, which are the number of training data pairs and the number of CNN parameters, respectively. While $\textsf{D}$ can be adjusted according to the amount of available data at the users, $P$ is usually unchanged during model training.  Here,  $P$ is computed as
		\begin{align}
		P = \underbrace{N_\mathrm{CL}(CN_\mathrm{SF} W_x W_y)}_{\mathrm{Convolutional\hspace{1pt} Layers}} +   \underbrace{\kappa N_\mathrm{SF}  W_x W_yN_\mathrm{FCL} \footnotesize }_{\mathrm{Fully\hspace{1pt} Connected\hspace{1pt} Layers}},
		\end{align} where  $N_\mathrm{CL}=3$ is the number of convolutional layers and $C=3$ is the number of spatial ``channels''.  $W_x=W_y=3$ are the 2-D kernel sizes. As a result, we have $P=600,192$.

		\section{Numerical Simulations}
		\label{sec:Sim}

		The goal of the simulations is to compare the performance of the proposed FL-based channel estimation approach with the state-of-the-art model-based channel estimation technique SF-CNN~\cite{deepCNN_ChannelEstimation} together with the MMSE and LS estimation in terms of normalized MSE (NMSE)~\cite{mimoDeepPrecoderDesign}, defined by $\mathrm{NMSE} = \frac{1}{J_T K} \sum_{i=1}^{J_T} \sum_{k=1}^{K}$ $ \frac{\| \boldsymbol{\Sigma}_{k} - \hat{\boldsymbol{\Sigma}}_{k,i} \|_\mathcal{F}^2}{\| \boldsymbol{\Sigma}_{k}  \|_\mathcal{F}^2   },$
		for $J_T = 100$ number of Monte Carlo trials. We also present the validation RMSE of the training process, for which the validation dataset includes $20\%$ of the whole dataset $\mathcal{D}$. 
		
		The local dataset of each user includes $N=100$ different channel realizations for $K=8$ users. The number of antennas at the BS and the IRS are $M=64$ and $L=64$, respectively. Also, we select $N_\mathrm{r}=5$ and the number of pilots is $\bar{M}=32$, unless stated otherwise. The location of each user is selected as $\phi_{k,n}\in \Phi_k$ and  $ {\varphi}_{k,n} \in \bar{\Psi}_k$, where $\Phi_k$ and $\bar{\Psi}_k$ are the equally-divided subregions of the angular domain $\Theta$, i.e.,  $\Theta =\bigcup_{k} \Phi_k = \bigcup_{k}\bar{\Psi}_k$, respectively. The pilot data are generated as  ${\mathbf{S}} = \mathbf{I}_{M}$ and $\overline{\mathbf{S}}$ is composed of the $\bar{M}$ columns of $\mathbf{S}$. During training, $N=200$ channel realizations are conducted and we have added AWGN on the input data for three SNR levels, i.e., SNR$=\{20, 25, 30\}$ dB,  for $G=160$ realizations in order to provide robust performance against noisy input~\cite{elbir_LIS}. As a result, the number of input-output pairs in the whole  training dataset is $\textsf{D}= 3KNG = 3\cdot8\cdot200\cdot160=768,000$. The proposed CNN model is realized and trained in MATLAB on a PC with a $2304$-core GPU. For CL, we use the SGD algorithm with momentum of $0.9$ and the mini-batch size $M_B = 128$,  and  update the network parameters with learning rate $0.001$. For FL, we train the CNN for $T=100$ iterations/rounds. Once the training is completed, the labels of the validation data (i.e., $20\%$ of the whole dataset) are used in prediction stage. During the prediction stage, each user estimates its own channel by feeding the CNN with $\boldsymbol{\Upsilon}_k$ and obtains  $\hat{\mathbf{h}}_{\mathrm{BS},k}$ and $\hat{\mathbf{G}}_k$ at the output.
		
		Let us first compare the transmission overhead of the proposed FL approach with CL. Using the analysis in Section~\ref{sec:DataComp} and the above system settings, we get $\mathcal{T}_\mathrm{CL} = 1.1\times 10^{10}$ whereas $\mathcal{T}_\mathrm{FL}=9.6\times 10^8$, which is approximately $12$ times lower than $\mathcal{T}_\mathrm{CL}$. This demonstrates the effectiveness of the FL over CL in terms of transmission overhead. The complexity of the proposed CNN approach is also lower than that of SF-CNN, for which the detailed complexity analysis is presented in~\cite{elbir2020_FL_CE}. In the sequel, we present the NMSE performance of the proposed approach.

		\begin{figure}[t]
			\centering
			\subfloat[]{\includegraphics[draft=false,width=\columnwidth]{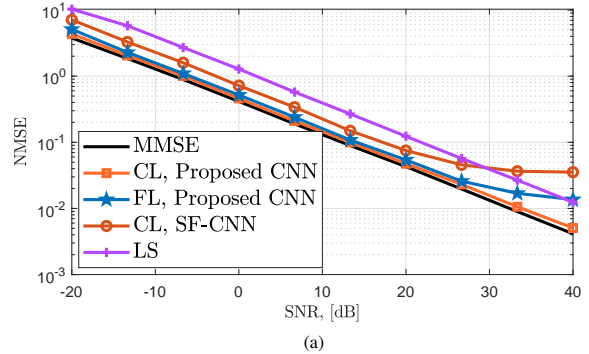}\label{fig_SNR} } 
			\\
			\subfloat[]{\includegraphics[draft=false,width=\columnwidth]{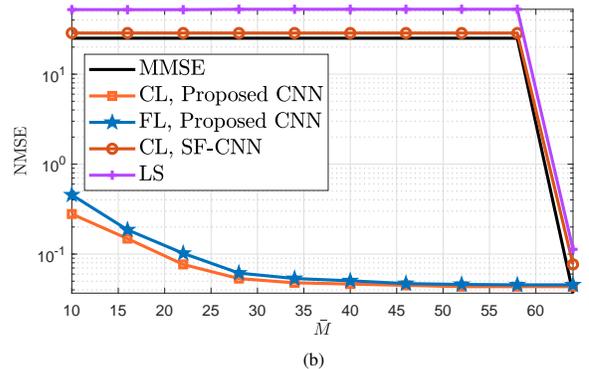} \label{fig_Number_of_pilots}} 
			\caption{ Channel estimation NMSE  with respect to (a) $\mathrm{SNR}$ when $\bar{M}=32$  and (b) $\bar{M}$ when $\mathrm{SNR} = 20$ dB.	}
			\label{fig_SNR_IRS}
		\end{figure}
		
		Fig.~\ref{fig_SNR} shows the channel estimation NMSE with respect to SNR. Notice that CL provides better performance than that of FL for the proposed CNN model since it has access to the whole dataset at once. Nevertheless, FL has satisfactory channel estimation performance despite decentralized training and  outperforms SF-CNN with CL. Specifically, the proposed CNN with FL and CL have similar NMSE for SNR$\leq 25$ dB and the performance of FL maxes out in high SNR regime. This is because the learning model loses  precision due to FL training and cannot perform better. SF-CNN also exhibits similar behavior but performs worse than the proposed method. This is because SF-CNN has  convolutional-only layers. In contrast, the proposed CNN includes both convolutional and fully connected layers, exhibiting better feature extraction and data mapping performance.
		
		Fig~\ref{fig_Number_of_pilots} shows the channel estimation NMSE with respect to the number of pilot signals $\bar{M}$ for SNR$=20$ dB. We can see that the proposed approach provides significantly better performance in the presence of insufficient pilot signals, i.e., $\bar{M} < M$, whereas the other algorithms perform poorly since they rely on the complete channel data, demanding $\bar{M}\geq M$.

		\section{Conclusions}
		\label{sec:Conc}
		We propose an FL framework for channel estimation in IRS-assisted massive MIMO systems. Via simulations, we demonstrate that FL-based channel estimation approach requires approximately $60\%$ fewer less pilot signals as compared to the model-based techniques. Moreover, FL-based training enjoys approximately $12$ times lower transmission overhead than  CL-based training while providing a channel estimation performance close to CL. We further observe that at least $15$ dB SNR on the model parameters is required for reliable channel estimation performance, i.e, $\mathrm{NMSE}\leq 0.01$. As a future work, we plan to study the FL-based models for joint channel estimation and IRS reflect beamformer design.

		\clearpage
		\balance
		\bibliographystyle{IEEEtran}
		\bibliography{IEEEabrv,references_070_journal}

	\end{document}